\documentclass[aps,prl,twocolumn,superscriptaddress]{revtex4-2}

\usepackage{graphicx}
\usepackage{amssymb}
\usepackage{amsmath}
\usepackage{bm}
\usepackage{xcolor}
\usepackage{braket}
\usepackage[normalem]{ulem}

\begin{document}
\title{Energy spectra and cascade in the spin turbulence of a driven spinor Bose-Einstein condensate}

\author{Junghoon Lee}
\affiliation{Department of Physics and Astronomy, Seoul National University, Seoul 08826, Korea}
\affiliation{NextQuantum, Seoul National University, Seoul 08826, Korea}

\author{Jongmin Kim}
\affiliation{Department of Physics and Astronomy, Seoul National University, Seoul 08826, Korea}
\affiliation{NextQuantum, Seoul National University, Seoul 08826, Korea}

\author{Donggyu Lee}
\affiliation{Department of Physics and Astronomy, Seoul National University, Seoul 08826, Korea}
\affiliation{NextQuantum, Seoul National University, Seoul 08826, Korea}

\author{Y. Shin}
\email{yishin@snu.ac.kr}
\affiliation{Department of Physics and Astronomy, Seoul National University, Seoul 08826, Korea}
\affiliation{NextQuantum, Seoul National University, Seoul 08826, Korea}
\affiliation{Institute of Applied Physics, Seoul National University, Seoul 08826, Korea}


\begin{abstract}
We investigate the spin-interaction energy spectrum of spin turbulence in a driven spinor Bose–Einstein condensate. Continuous spin driving of a spin-1 condensate produces a nonequilibrium steady state with spatially fluctuating magnetization. We observe a power-law scaling consistent with the $-7/3$ exponent predicted for spin-wave turbulence, which persists across our full range of drive strengths despite substantial changes in the spectral anisotropy. After switching off the drive, we track the free-decay evolution and find evidence consistent with a direct cascade of spin-interaction energy toward higher wavenumbers. These results establish an energy-spectral hallmark of spin turbulence and enable quantitative studies of cascade dynamics in spinor superfluids.
\end{abstract}

\maketitle
Turbulence is a ubiquitous yet still enigmatic phenomenon, not only due to its complex spatiotemporal structures but also because it exemplifies nonequilibrium transport: nonlinear interactions redistribute conserved quantities across scales~\cite{pope2000turbulent,eyink2006onsager}. In classical fluids, this is captured by the energy cascade and inertial-range scaling, typified by the Kolmogorov $-5/3$ power law~\cite{kolmogorov1991local,alexakis2018cascades}. Quantum fluids provide a complementary setting with a well-defined order parameter, quantized circulation, and tunable dissipation~\cite{barenghi2014introduction,tsatsos2016quantum}. Quantum turbulence has been explored in superfluid helium~\cite{vinen2002quantum,skrbek2021phenomenology}, atomic Bose–Einstein condensates (BECs)~\cite{henn2009emergence,bradley2012energy,kwon2014relaxation,navon2016emergence,gauthier2019giant,johnstone2019evolution,dogra2023universal,zhao2025kolmogorov}, and polariton condensates~\cite{amo2011polariton,panico2023onset}, revealing both classical-like phenomenology and distinctly quantum routes to cascade dynamics. 

A qualitatively new regime arises when the superfluid possesses internal spin degrees of freedom, as in superfluid $^3$He~\cite{Salomaa1987} and spinor BECs~\cite{ohmi1998bose,kawaguchi2012spinor}. In these systems, the mass flow is intrinsically coupled to the magnetic order, meaning that turbulence involves multiple interacting hydrodynamic channels. Nonlinear interactions among superflows, spin waves, and spin textures can produce turbulent states in which the spin and mass transport coevolve, governed by symmetries and conservation laws absent in scalar superfluids.
In typical atomic spinor BECs, the energy scale of spin excitations lies well below that of phonons, allowing cascade processes to develop largely within the spin sector~\cite{yukawa2012hydrodynamic,fujimoto2016direct,karl2013tuning,fujimoto2013spin}. This regime, known as {\it spin turbulence}, offers a platform for studying far-from-equilibrium transport phenomena in magnetic media.
Experimentally, turbulent spin dynamics has been observed in various settings, i.e., after quantum quenches~\cite{sadler2006spontaneous,bookjans2011quantum,kang2017emergence,symes2018dynamics,huh2024,kang2020crossover}, under strong spin currents~\cite{Takeuchi2010,fujimoto2012counterflow,kim2017critical}, and with continuous spin driving~\cite{hong2023spin,kim2024chaos,Lee2025}, producing complex spin textures. However, direct measurements of a spin-energy spectrum demonstrating a spin-sector cascade have remained elusive.

In this paper, we report measurements of the spin-interaction energy spectrum in a turbulent spin-1 BEC and identify a robust scaling regime with a distinct spectral signature. By establishing a nonequilibrium steady state through continuous spin driving and measuring spatial magnetization profiles along all spin axes, we reconstruct the spin-interaction energy spectrum. The spectrum shows power-law scaling with an exponent close to the theoretical value of $-7/3$ for spin-wave turbulence~\cite{fujimoto2013spin,fujimoto2016direct}.
In complementary measurements of freely decaying turbulence, we find evidence consistent with the direct transfer of the spin-interaction energy toward higher wavenumbers. These results establish a key spectroscopic signature of spin turbulence and provide a rationale for quantitative tests of universality and coupled-cascade dynamics in spinor superfluids.

The order parameter of a spin-1 BEC is expressed as $\Psi(\mathbf{r}) = \sqrt{n(\mathbf{r})}\,e^{i\phi(\mathbf{r})}\boldsymbol{\zeta}(\mathbf{r})$, where $n$ denotes the condensate density, $\phi$ is the superfluid phase, and $\boldsymbol{\zeta} = (\zeta_{+1}, \zeta_0, \zeta_{-1})^\text{T}$ represents the normalized spinor representing the spin state. Spatial variations of $\phi$ and $\boldsymbol{\zeta}$ give rise to coupled mass and spin currents~\cite{yukawa2012hydrodynamic}, and their hydrodynamic velocities $\mathbf{u}$ and $\mathbf{v}_i$ ($i=x,y,z$) are expressed, respectively, as
\begin{align}
\mathbf{u}(\mathbf{r}) &= \frac{\hbar}{m}\left(\nabla \phi - i\boldsymbol{\zeta}^\dagger\nabla\boldsymbol{\zeta}\right) \\
\mathbf{v}_i(\mathbf{r}) &= 
\frac{\hbar}{m}\left(f_i\nabla\phi
-\frac{i}{2}
\left[\boldsymbol \zeta^\dagger F_i\nabla \boldsymbol\zeta-(\nabla \boldsymbol\zeta^\dagger)F_i\boldsymbol \zeta \right] \right),
\end{align}
where $\boldsymbol{F}=(F_x,F_y,F_z)$ is the vector of the spin-1 matrices and $\mathbf{f} = (\textrm{f}_x,\textrm{f}_y,\textrm{f}_z)$ with $\textrm{f}_i = \boldsymbol{\zeta}^\dagger {F}_i \boldsymbol{\zeta}$ is the spin vector.
In a nonequilibrium turbulent regime, fields $\mathbf{u}$, $\mathbf{v}_i$, and $\mathbf{f}$ become strongly disordered in space, giving rise to highly complex spin-mass hydrodynamics. 

Besides the kinetic energies of mass and spin currents, a spinor condensate also stores interaction energy in its magnetization,
\begin{equation}
\mathcal{E}_s=\int \frac{1}{2} c_2 |\mathbf{M}(\mathbf{r})|^2 d^3r,
\end{equation}
with $\mathbf{M}(\mathbf{r}) \equiv n(\mathbf{r}) \mathbf{f}(\mathbf{r})$.
This spin-interaction energy underlies the nonlinear spin dynamics and provides a natural quantity for characterizing spin turbulence. 
Hydrodynamic theory predicts universal power laws in the spin-interaction energy spectrum for small spin magnitude~\cite{fujimoto2013spin}: a $k^{-7/3}$ scaling at high wavenumbers where convective transfer of spin-interaction energy dominates, and a $k^{-1}$ scaling at low wavenumbers where the local spin precession from spin-torque coupling prevails~\cite{yukawa2012hydrodynamic, fujimoto2013spin}.
The focus of this work is to investigate the power spectrum of the magnetization field in turbulent spinor BECs.

\begin{figure}[t]
\includegraphics[width=8.6cm]{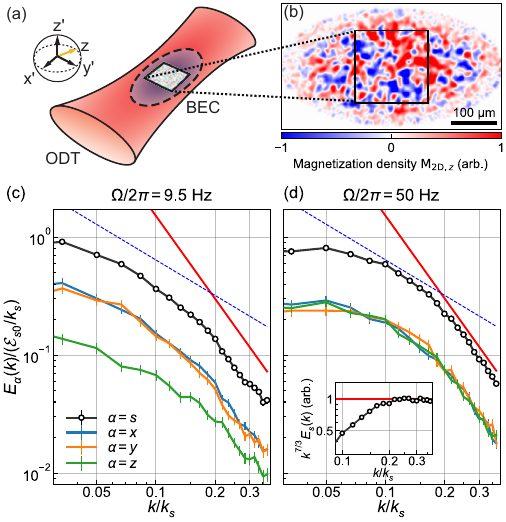}
\centering
\caption{
Spin turbulence in a trapped spinor Bose-Einstein condensate (BEC).
(a) Experimental schematic. An $F=1$ $^{23}$Na BEC in an optical dipole trap (ODT) is continuously driven by a transverse resonant RF field with Rabi frequency $\Omega$. The static magnetic field defines the $z$ axis, with $\hat{z}\simeq0.7(-\hat{x}'+\hat{y}')+0.1\hat{z}'$ in the laboratory frame $(x',y',z')$.
(b) Representative image of the effective areal magnetization $\textrm{M}_{\mathrm{2D},z}$ in the turbulent steady state. The box indicates the region used for spectral analysis.
(c,d) Spin-interaction energy spectra for (c) $\Omega/2\pi=9.5~\mathrm{Hz}$ and (d) $50~\mathrm{Hz}$. Colored lines show $E_i(k)$ ($i=x,y,z$), with the transverse $x$ and $y$ axes defined by the RF phase in the rotating frame; the black line shows the total spectrum $E_s(k)$. Spectra are normalized by $\mathcal{E}_{s0}/k_s$, where $\mathcal{E}_{s0}$ is the fully magnetized spin-interaction energy and $k_s=2\pi/\xi_s$ with $\xi_s$ being the spin healing length (see the text). Error bars denote one standard error of the mean~\cite{SM}. Blue dashed and red solid lines indicate $k^{-1}$ and $k^{-7/3}$ scalings, respectively.
The inset shows the compensated spectrum $k^{7/3} E_s(k)$.
}
\label{fig1}
\end{figure}

\begin{figure*}[t]
\includegraphics[width=18.0cm]{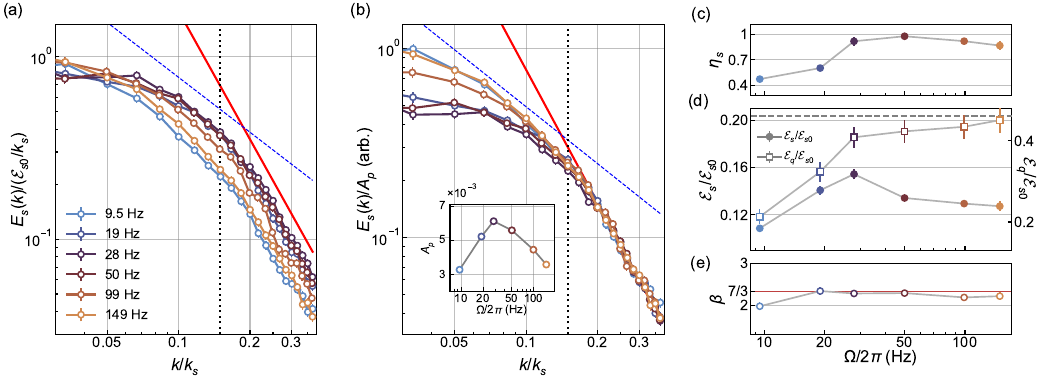}
\centering
\caption{Spin-interaction energy spectra in spin turbulence.
(a) Energy spectra $E_s(k)$ at various driving strengths $\Omega$. The vertical dotted line indicates $\bar k_\ell$, the mean lower bound of the high-$k$ scaling range across all $\Omega$ values~\cite{SM}.
(b) Rescaled energy spectra $E_s(k)/A_p$. The prefactor $A_p$ was extracted by fitting $E_s(k)/(\mathcal{E}_{s0}/k_s) = A_p(k/k_s)^{-7/3}$ to each spectrum in the range $k \geq 1.3\,\bar k_\ell$. The inset shows $A_p$ as a function of $\Omega$ (see Supplemental Material~\cite{SM} for a model describing the energy injection rate).
(c)–(e) Turbulence parameters as a function of $\Omega$: (c) Energy anisotropy $2\mathcal{E}_{z}/\mathcal{E}_{\perp}$, where $\mathcal{E}_i=\int E_i(k) dk$ and $\mathcal{E}_\perp=\mathcal{E}_x+\mathcal{E}_y$; (d) total spin-interaction energy $\mathcal{E}_s=\sum_i \mathcal{E}_i$ and quadratic Zeeman energy $\mathcal{E}_q$; and (e) power-law exponent $\beta$, determined from the fitting of $E_s(k)=Ak^{-\beta}$ to the high-$k$ scaling region ($k \geq 1.3\,\bar k_\ell$). The horizontal dashed line in (d) denotes the $\mathcal{E}_q$ value for an equal mixture of the three spin components. Error bars in (c) and (d) represent the propagated $1\sigma$ standard error from the spectral integration, while those in (e) denote the $1\sigma$ standard error of the weighted power-law fits.
}
\label{fig2}
\end{figure*}

Our experiment starts with the preparation of a BEC of $F=1$ $^{23}$Na atoms in a spin-independent optical dipole trap.
The trap is highly oblate, with trapping frequencies $(\omega_{{x}'},\omega_{{y}'},\omega_{{z}'})=2\pi\times(4.4, 8.9, 420)$ Hz.
A uniform magnetic field is applied along the $z$ axis [Fig.~\ref{fig1}(a)] and produces a quadratic Zeeman shift $q/h = 10.5$~Hz.
For antiferromagnetic interactions ($c_2>0$), the spin-interaction energy is minimized for unmagnetized polar states with $\mathbf{f}=0$. The quadratic Zeeman shift introduces spin anisotropy, and for $q>0$, the magnetic ground state is the axial polar state $\boldsymbol{\zeta} = (0,1,0)^\text{T}$.
The characteristic spin-interaction energy per atom is $\varepsilon_{s}=c_2 {n}_0\simeq h\times 50$~Hz, where ${n}_0$ is the peak condensate density.
In the highly oblate geometry with tight confinement ($\hbar\omega_{z'}>\varepsilon_{s}$), spin excitations along the ${z}'$ axis are strongly suppressed such that the condensate has an effectively two-dimensional spin texture.

To generate spin turbulence, we apply a continuous resonant radio-frequency (RF) magnetic field with Rabi frequency $\Omega$ to rotate the spin, as demonstrated in \cite{hong2023spin,kim2024chaos}.
Briefly, this continuous spin driving method is based on the spin-energy anisotropy of the system due to the quadratic Zeeman shift. 
By rotating the spin state at a controlled rate, energy is injected in the form of quadratic Zeeman energy, and is subsequently converted dynamically into spin-wave excitations via the spin-mixing process~\cite{kawaguchi2012spinor}, producing a turbulent state with a spatially disordered spin texture~\cite{SM}.

After holding the system for $2$~s under RF driving to ensure steady turbulence, we measure the \textit{in situ} column density distributions $\bar{n}_{m_F}(x',y')$ of the $m_F=\pm1$ components projected onto spin axis $i$, using a multiple imaging technique~\cite{huh2024,SM}. 
The measurement axis $i=x,y,z$ is selected by applying short $\pi/2$ RF pulses immediately before imaging to rotate the spin basis into the imaging direction. 
For the 2D spin texture with $\mathbf{f}(x',y')$, the spin-interaction energy can be written as $\mathcal{E}_s=
\frac{c_2}{2}\int |\mathbf{M}_\text{2D}|^2 dx' dy'$, where $\mathbf{M}_\text{2D}(x',y')$ is the effective areal magnetization density and its components are given by $\textrm{M}_{\text{2D},i}= \lambda(\bar{n}_{+1}-\bar{n}_{-1})$, with $\lambda(x',y')=[\int n^2 dz']^{1/2}/ \int n dz'$ (see Supplemental Material~\cite{SM} for details). We obtain $\textrm{M}_{\text{2D},i}(x',y')$ from the measured column densities and using the Thomas-Fermi (TF) approximation with $\lambda^2=3/(5 z'_{R})$ and $z'^{2}_{R}=R_{z'}^2(1-\frac{x'^2}{R_{x'}^2}-\frac{y'^2}{R_{y'}^2})$ [Fig.~\ref{fig1}(b)]. Here, $R_j$ denotes the TF radius along the $j$ direction and $R_{z'}=(\frac{\omega_{x'}\omega_{y'}}{\omega_{z'}^2} R_{x'} R_{y'})^{1/2}$. $R_{x',y'}$ were determined from the TF profile fit to the measured total column density distribution.

The one-dimensional spin-interaction energy spectrum is calculated from $\mathbf{M}_\text{2D}(x',y')$ using the equations, 
\begin{eqnarray}
\label{energyequation}
E_s(k) &=& \sum_i E_{i}(k) \nonumber \\ 
E_{i}(k) &=& \frac{c_2}{8\pi^2} \int_0^{2\pi} d\varphi_k k |\widetilde{{\textrm M}}_{\text{2D},i}(\mathbf{k})|^2.
\end{eqnarray}
Here, $\widetilde{{\textrm M}}_{\text{2D},i}(\mathbf{k})$ is the two-dimensional Fourier transform of ${\textrm M}_{\text{2D},i}({x',y'})$, $k=|\mathbf{k}|$, and $\varphi_k$ is the azimuthal angle in $\mathbf{k}$-space (see Supplemental Material~\cite{SM} for details).
To minimize density inhomogeneity effects,  we evaluate $E_s(k)$ within a central 160~$\mu\text{m}\times 160~\mu\text{m}$ square region of the trapped sample, where the column density varies by less than 25\% [Fig.~\ref{fig1}(b)]. 
In our experiment, the spin-interaction energy for a fully magnetized condensate is estimated to be $\mathcal{E}_{s0}\approx 0.35  \varepsilon_s N$ with $N$ being the condensate atom number in the region of interest. 
The effective spin healing length for the quasi-2D condensate is given by $\xi_s=\sqrt{\frac{\hbar^2}{2m (2\mathcal{E}_{s0}/N)}} \approx 2.6~\mu\text{m}$ with the reciprocal length $k_s=2\pi/\xi_s$. 

In Figs.~\ref{fig1}(c) and (d), we show two sets of energy spectra $E_i(k)$ for different driving strengths $\Omega$.
For a weak driving case ($\Omega/2\pi \approx 10$ Hz), the spectra exhibit axis-symmetric anisotropy with $E_{x}\approx E_{y}>E_{z}$. The suppression of $E_z$ reflects the intrinsic spin anisotropy of the system with the quadratic Zeeman shift $q>0$, which energetically penalizes longitudinal magnetization along $z$.
As the driving strength increases ($\Omega/2\pi = 50$ Hz), the spectra measured along all spin directions become identical, indicating the emergence of an isotropic spin-turbulent regime~\cite{hong2023spin}. In this regime, rapid spin rotation couples transverse and longitudinal magnetization strongly enough to overcome the system’s intrinsic anisotropy. 

The total energy spectrum $E_s(k)$ shows power-law scaling at high wavenumbers. The inset of Fig.~\ref{fig1}(d) shows the compensated spectrum $k^{7/3} E_s(k)$ for $\Omega/2\pi = 50$~Hz, which exhibits a clear plateau, consistent with the predicted $k^{-7/3}$ scaling in the high-$k$ range~\cite{fujimoto2013spin}.
Figure~\ref{fig2}(a) shows $E_s(k)$ for driving strengths up to $\Omega/2\pi \approx 150$~Hz.
The high-$k$ power-law regime is largely insensitive to $\Omega$: it emerges near $\bar k_\ell\simeq 0.15k_s$~\cite{SM} and extends to the imaging resolution limit of $k_\text{res} \simeq 0.38k_s$. 
To highlight the universality of this behavior, Fig.~\ref{fig2}(b) displays the rescaled spectra $E_s(k)/A_{p}$, where $A_{p}$ is obtained by fitting $E_s(k)/(\mathcal{E}_{s0}/k_s) = A_{p}\,(k/k_s)^{-7/3}$ to each spectrum over a fixed window of $1.3\bar k_\ell \leq k \leq k_\mathrm{res}$. 
The resulting collapse demonstrates the robustness of the $k^{-7/3}$ scaling across the full range of driving strengths~\cite{footnote3}.
At low wavenumbers, the spectra exhibit a weak tendency toward $k^{-1}$-like behavior over a limited interval, although the available spectral range is too narrow to establish a robust power-law scaling.

\begin{figure*}[t]
\includegraphics[width=17.8cm]{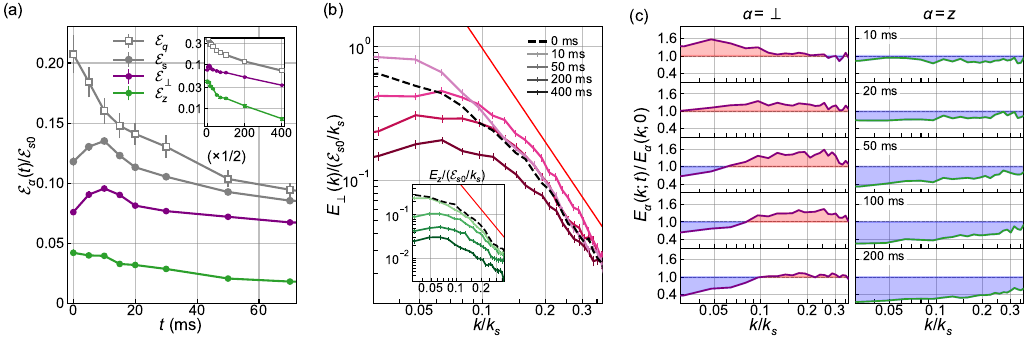}
\centering
\caption{Free decay of spin turbulence.
(a) Time evolution of the quadratic Zeeman energy $\mathcal{E}_q$ (gray squares), total spin-interaction energy $\mathcal{E}_s$ (gray circles), and its transverse ($\mathcal{E}_\perp=\mathcal{E}_x+\mathcal{E}_y$, purple), and longitudinal ($\mathcal{E}_z$, green) components. 
(b) Transverse spectra $E_{\perp}(k)=E_{x}(k)+E_{y}(k)$ for hold times $t=0, 10, 50, 200, 400$-ms after switching off the drive. The inset shows the corresponding longitudinal spectra $E_{z}(k)$. Red solid lines indicate the $k^{-7/3}$ power law, and black dashed lines denote the spectra at $t=0$. 
(c) Spectral ratio ${E}_{\alpha}(k,t)/{E}_{\alpha}(k,0)$ for the transverse ($\alpha=\perp$, left) and longitudinal ($\alpha=z$, right) components at different hold times. Red (blue) shading indicates spectral enhancement (depletion) relative to $t=0$.
Error bars in (a) and (b) denote the standard error of the mean from repeated measurements.
}
\label{fig3}
\end{figure*}

Figures~\ref{fig2}(c)-\ref{fig2}(e) summarize how the steady turbulent state evolves with the drive strength $\Omega$. We quantify the spectral anisotropy using  $\eta_s\equiv 2\mathcal{E}_{z}/\mathcal{E}_\perp$, where $\mathcal{E}_{i}=\int E_{i}(k)dk$~\cite{SM} and $\mathcal{E}_\perp=\mathcal{E}_x+\mathcal{E}_y$. As $\Omega$ increases, $\eta_s$ increases and approaches unity at $\Omega_{\mathrm{iso}}/2\pi \simeq 30$~Hz, indicating near-isotropic turbulence.
Notably, the total spin-interaction energy $\mathcal{E}_s=\mathcal{E}_\perp+\mathcal{E}_z$ peaks around this isotropization point. For $\Omega>\Omega_{\mathrm{iso}}$, $\mathcal{E}_s$ decreases despite the fact that the quadratic Zeeman energy $\mathcal{E}_q$ continues to increase as the spin composition approaches an equal mixture of the three spin components~\cite{hong2023spin}. Here, $\mathcal{E}_q=q N (1-\rho_{0})$ with $\rho_0$ being the $m_F=0$ population fraction for $z$ axis.
This anticorrelation at high $\Omega$ suggests that rapid spin rotation outpaces the intrinsic spin-exchange timescale for generating magnetization, in turn suppressing the spin-interaction energy growth. 
Consistent with the decrease in $\mathcal{E}_s$, Ref.~\cite{hong2023spin} reported reduced heating at a high $\Omega$.
For $\Omega/2\pi > 100$ Hz, the steady turbulent state becomes fully isotropic in terms of both the spin composition and the spin-interaction energy spectrum.
We emphasize that the spectrum $E_s(k)$ retains the $-7/3$ power-law scaling across our full $\Omega$ range [Fig.~\ref{fig2}(e)], despite the fact that the spectral anisotropy changes substantially [Fig.~\ref{fig2}(c)].

To elucidate how the spin-interaction energy is redistributed across $k$-space and among the spin components, next we investigate freely decaying turbulence. 
In this case, first we prepare a fully spin-isotropic, turbulent steady state at $\Omega/2\pi\approx 150$~Hz, after which we track the subsequent spectral evolution after switching off the RF drive.

Figure~\ref{fig3}(a) shows the time evolution of the quadratic Zeeman energy and spin-interaction energies $\mathcal{E}_\alpha$ ($\alpha=q,s,\perp, z$). Immediately after switching off, $\mathcal{E}_s$ exhibits a rapid overshoot, occurring predominantly in the transverse sector, while $\mathcal{E}_q$ decreases monotonically.
The initial growth rate of $\mathcal{E}_s$ constitutes a substantial fraction of the decay rate of $\mathcal{E}_q$, demonstrating that the energy injection into the spin-interaction energy sector is fueled by the release of the quadratic Zeeman energy~\cite{footnote2}.
After the initial surge, all energy components decay as the system relaxes to the polar ground state, while the spin fraction $\rho_0$ increases.
During this relaxation step, $\mathcal{E}_z$ decays nearly twice as rapidly as $\mathcal{E}_\perp$~\cite{footnote1} (Fig.~\ref{fig3}(a) inset), which is consistent with the anisotropic steady turbulence with $\eta_s<1$ observed for small $\Omega$.

The energy surge in $\mathcal{E}_\perp$ lasts for about $10$~ms before $\mathcal{E}_\perp$ begins to decay. 
This time span corresponds to a quarter period of the spin rotation at $\Omega_{\mathrm{iso}}/2\pi \approx 30$~Hz, the crossover frequency for spin isotropization, at which the steady-state spin-interaction energy is maximized. 
This comparison corroborates the interpretation that sufficiently rapid spin rotation interrupts the buildup of transverse ferromagnetic correlations, while simultaneously promoting mixing between longitudinal and transverse magnetization.
Consistently, for a uniform transverse-polar state $\boldsymbol{\zeta} = \frac{1}{\sqrt{2}}(1,0,1)^\text{T}$, the characteristic frequency for magnetization generation is estimated as $\sqrt{q(2\varepsilon_s-q)}/h \approx 30$~Hz~\cite{kang2017emergence,symes2018dynamics,kang2020crossover}.
We note that for small $\Omega<\Omega_\text{iso}$, no discernible energy surge was observed in the decaying turbulence~\cite{SM}.

The corresponding spectral evolution during free decay is shown in Fig.~\ref{fig3}(b). The incipient injection is concentrated at low $k\lesssim 0.1k_s$ in the transverse spectrum $E_{\perp}(k)=E_x(k)+E_y(k)$. 
After this buildup, the low-$k$ spectral weight decays rapidly, while high-$k$ modes show delayed growth and then decay.
Figure~\ref{fig3}(c) shows the normalized spectral evolution $E_{\perp}(k;t)/E_{\perp}(k;0)$, for several hold times $t$, highlighting the delayed high-$k$ response.
We interpret this temporal separation as inter-scale transport: energy injected at large scales requires a finite time to reach smaller scales, consistent with a direct cascade.
The $k^{-7/3}$ scaling for $k>\bar k_\ell$ persists up to 200~ms~\cite{SM}.
In the longitudinal sector, there is no pronounced low-$k$ injection, but high-$k$ modes still decay more slowly than low-$k$ modes, implying a cascade-like redistribution within the spin sector.

The observation of an energy cascade with power-law scaling is suggestive of an inertial range with an approximately constant energy flux~\cite{gauthier2019giant,johnstone2019evolution,dogra2023universal,alexakis2018cascades,fujimoto2013spin}, but our current data do not yet fully resolve this.
In decaying turbulence, the net gain at a high $k$ accounts for only part of the transient excess accumulated at a low $k$.
This implies that a substantial fraction of the injected energy is likely transferred to channels not captured by $E_s(k)$, such as the spin-texture energy or the kinetic energy of the mass flow~\cite{yukawa2012hydrodynamic}.
Such inter-channel coupling is intrinsic to multicomponent superfluids, and quantifying the transfers among these coupled reservoirs will be essential when attempting to devise a complete description of energy transport in spin turbulence.
We also note that $k^{-7/3}$ scaling has been discussed in multiple contexts, including spin turbulence involving domain walls and vortices~\cite{fujimoto2013spin,fujimoto2012counterflow,jung2023random} and weak spin-wave turbulence mediated by resonant spin-wave interactions~\cite{fujimoto2016direct}. 
The spectral exponent may not uniquely identify the underlying transport process~\cite{rosenhaus2024interaction,liu2025degrees}, an issue that merits further investigation.

Finally, we comment on the energy injection mechanism under continuous spin driving.
Given that the driving is implemented via global spin rotation, it is expected that energy injection is concentrated mainly at a low-$k$ region, as observed in the decaying turbulence case. However, we currently lack a quantitative framework for determining the low-$k$ injection scale, particularly in the turbulent steady state.
Because magnetization generation depends on the instantaneous spatial structure of the spin texture, its interplay with driven rotation should determine the injection rate and, together with dissipation, the steady value of $\mathcal{E}_s$~\cite{SM}. Clarifying this feedback mechanism is an important direction for future work.
Furthermore, chaotic spin dynamics under continuous driving was shown to be essential for sustaining the turbulent state~\cite{kim2024chaos}, and how this chaos governs energy flows and inter-scale transfers in the steady state remains another intriguing open question. 

In summary, we have demonstrated that a driven spin-1 condensate can sustain turbulence with a spin-interaction energy spectrum consistent with $-7/3$ power-law scaling and that the spin-interaction energy is injected through transverse magnetization generation and subsequently transferred from large to small scales. 
Using the tunability of $\Omega$ and $q$, this platform can be extended to access different regimes of turbulence.
Varying $\Omega$ enables systematic studies of anisotropic turbulence in comparison with the corresponding classical counterparts~\cite{biferale2005anisotropy}, and adjusting $q$, especially to negative values that alter the spin-rotation symmetry~\cite{bookjans2011quantum,kang2017emergence}, may offer a means by which to probe how the hierarchy of spin-energy channels reorganizes the cascade dynamics.

\begin{acknowledgments}
We thank Junhwan Kwon for insightful discussions. This work was supported by the National Research Foundation of Korea (Grant Nos. RS-2023-NR077280, RS-2023-NR119928, and RS-2024-00413957).
\end{acknowledgments}

\clearpage
\onecolumngrid

\begin{center}
{\large \bf Supplemental Material}
\end{center}

\setcounter{equation}{0}
\setcounter{figure}{0}
\setcounter{table}{0}

\makeatletter
\renewcommand{\theequation}{S\arabic{equation}}
\renewcommand{\thefigure}{S\arabic{figure}}
\renewcommand{\thetable}{S\arabic{table}}
\makeatother

\subsection{Methods}
\subsubsection{Sample preparation}

We prepared a spinor Bose-Einstein condensate (BEC) of $\approx 1.2 \times 10^7$ $^{23}$Na atoms in the $|F=1, m_F=0\rangle$ state in an optical dipole trap (ODT), with a thermal fraction below 10\%. 
The ODT was highly oblate with trapping frequencies $\{ \omega_{x'}, \omega_{y'},\omega_{z'}\} = 2\pi \times \{4.4, 8.9, 420\}$~Hz.
A bias magnetic field $B_0=0.20$~G set the Larmor frequency to $\omega_0/2\pi \approx 140$~kHz and the quadratic Zeeman shift to $q/h = 10.5$~Hz.
The peak atomic density in the trapped condensate was $n_0 \approx 8.7(4) \times 10^{19}~\text{m}^{-3}$, corresponding to a chemical potential $\mu = h \times 1.4(1)$~kHz and a characteristic spin-interaction energy $\varepsilon_s=c_2 n_0 = h\times 48(2)$~Hz. 

\subsubsection{Spin driving and turbulence generation}
\label{sec:spin_driving}

We generated turbulence using resonant RF spin driving~\cite{hong2023spin}.
The RF magnetic field was applied transverse to the static bias field for 2~s, with Rabi frequencies $\Omega/2\pi \approx 10$-$150$~Hz.
Using the rotating-wave approximation, in the frame rotating at the same frequency $\omega_{\mathrm{RF}}$ with the RF driving field, the mean-field Hamiltonian for the local spin dynamics is given by 
\begin{equation}
\mathcal{H}_{\mathrm{spin}}
= \hbar\delta F_z-\hbar\Omega F_x
+qF_z^2+c_2n\,\mathbf{f}\cdot\mathbf{F},
\label{eq:spin_hamiltonian}
\end{equation}
where $\delta=\omega_{\mathrm{RF}}-\omega_0$, $\mathbf{F}=(F_x,F_y,F_z)$ denotes the spin-1 matrices, $\boldsymbol{\zeta}$ is the normalized spinor, $\mathbf{f}=\boldsymbol{\zeta}^{\dagger}\mathbf{F}\boldsymbol{\zeta}$ is the spin vector per particle, and $n$ is the local atomic density.
The spin $z$ axis is along the static bias field $B_0$, and the transverse $x$ axis is set by the phase of the RF field.
In the resonant driving, $\delta=0$. Bias magnetic field fluctuations were identified at a level of about $1$~mG, which was found to enhance the chaoticity of the driven spin dynamics and facilitate near-complete randomization of the spin state~\cite{kim2024chaos, jung2023random}.

\subsubsection{Spin-selective imaging}

To probe the spin texture, we performed sequential absorption imaging of the atoms in the three Zeeman sublevels of the $F=1$ manifold, either immediately after switching off the RF field for steady-state measurements or after a variable hold time for decay measurements in the main text. For each $\ket{F{=}1,m_F}$ component, a short resonant microwave pulse was applied to transfer $\simeq30\%$ of the population to the corresponding $\ket{F{=}2,m_F'}$ state, which was then imaged on the $F{=}2\rightarrow F'{=}3$ cycling transition. 
The microwave frequency was tuned to the corresponding $\ket{F{=}1,m_F=-1,+1,0}\rightarrow\ket{F{=}2,m_F'=-2,+2,0}$ transitions. The three images were acquired sequentially with an interframe delay of $\Delta t\simeq1.1~\mathrm{ms}$. Under our imaging condition, off-resonant scattering of the probe light by $F{=}1$ atoms is negligible ($\lesssim10^{-3}$ photons per atom per pulse).

The interframe delay here was much shorter than the spin-interaction timescale, $h/\varepsilon_s\sim20~\mathrm{ms}$, to preserve the spin texture during the imaging sequence. The density redistribution was also expected to be negligible on the length scales analyzed: column-averaged sound speed $c_{s}=\sqrt{2\mu/3m}\approx4~\mathrm{mm/s}$ implies a propagation distance of $\lesssim4~\mu\mathrm{m}$ over $\Delta t$, a value smaller than our imaging resolution of $\approx6.8~\mu\mathrm{m}$. We further checked for imaging-induced perturbations by (i) taking consecutive images of the same spin component using the identical sequence, which showed negligible changes on the level of the shot noise, and (ii) permuting the imaging order, which produced no measurable differences in the reconstructed magnetization profiles.

In the second and third frames, atoms transferred to $F{=}2$ in previous steps appear as a diffuse background with an approximately Gaussian profile, resulting from photon recoil during earlier imaging pulses. We characterized this background by omitting the microwave transfer pulse for the subsequent frame; what remained was well described by a two-dimensional Gaussian. We then fit the tails outside the condensate region and subtracted the corresponding Gaussian background prior to reconstructing the spin-component densities.

Finally, we calibrated the relative optical densities of the three $m_F$ components using reference samples with well-defined spin compositions, in this case the easy-axis polar state with $(\rho_{+1},\rho_0,\rho_{-1})=(0,1,0)$ and a tilted polar state with $\approx(1,1,1)/3$. We verified that the total column density distribution reconstructed from the three images follows a Thomas-Fermi (TF) profile.

\subsubsection{Modulation transfer function calibration}
\label{sec:mtf_calibration}

The squared modulation transfer function (MTF), $|\mathrm{MTF}(\mathbf{k})|^2$, was calibrated from the Fourier power spectrum of in-situ absorption images of thermal clouds, following Ref.~\cite{hung2011extracting}. Thermal-gas images were acquired under optical and camera settings identical to those used in the main experiments. For each image, we computed the two-dimensional Fourier power spectrum and removed a constant offset estimated from the outermost Fourier-space border pixels, where the atomic signal is strongly suppressed by the finite imaging resolution. The background-subtracted spectra were then averaged over $100$ realizations.
The thermal clouds were sufficiently hot such that the thermal de Broglie wavelength was shorter than the imaging resolution, meaning that density fluctuations can be expected to be spatially white over the spatial-frequency range of interest. This "white-noise" input allows the measured power spectrum to be interpreted as $|\mathrm{MTF}(\mathbf{k})|^2$ up to the overall scale factor.

To parameterize $|\mathrm{MTF}(\mathbf{k})|^2$, we adopt the exit-pupil ansatz from Ref.~\cite{hung2011extracting}, and model
\begin{eqnarray}
|\mathrm{MTF}(\mathbf{k})|^2 &=& \exp(-2\rho^2/\tau^2)\,
\cos^2\!\bigl[\delta_\varphi + W(\rho,\vartheta)\bigr],
\label{eq:mtf_sq}\\
W(\rho,\vartheta) &=& A_s\rho^4 + A_a\rho^2\cos\!\bigl[2(\vartheta-\varphi)\bigr] + A_b\rho^2.
\end{eqnarray}
Here $\rho=|\mathbf{k}|/a$ is the normalized radial coordinate, $\delta_\varphi$ is an effective detection phase set primarily by the imaging detuning, and $W(\rho,\vartheta)$ represents a low-order wavefront aberration, including primary spherical aberration, astigmatism along axis $\varphi$, and defocus effects. We fixed $a$ to the highest spatial frequency allowed by the camera sampling (Nyquist limit) and determined the remaining parameters by a weighted nonlinear least-squares fit to the measured two-dimensional thermal-noise spectrum. In the fit, we excluded the central low-$|\mathbf{k}|$ pixels (up to the fourth pixel), which are affected by the finite cloud size and trapping profile, and we iteratively rejected outliers that deviate from the fitted curve by more than $3\sigma$.
The resolution cutoff wavenumber $k_{\mathrm{res}}$ was determined for the signal-to-noise ratio $\mathrm{SNR}>1$, which yielded $k_{\mathrm{res}}\simeq 0.92~\mu\mathrm{m}^{-1}$.

\subsection{Determination of spin-interaction energy spectra}

For a given spin axis $i\in\{x,y,z\}$, we measure the \textit{in situ} column-density distributions $\bar n_{m_F}(x',y')$ of the $m_F=\pm1,0$ components. From these, we obtain the column magnetization density $\bar M_i(x',y')=\bar n_{+1}-\bar n_{-1}$ and the total column density $\bar n(x',y')=\sum_{m_F}\bar n_{m_F}$. For our highly oblate sample, we treat the spin texture as effectively two dimensional and write $\mathbf f(x',y')=\bar{\mathbf M}/\bar n$, where $\bar{\mathbf M}=(\bar M_x,\bar M_y,\bar M_z)$.
Then, the spin-interaction energy is expressed as
\begin{eqnarray}
\mathcal E_s
&=&\int \frac{c_2}{2}\,|n(\mathbf r')\mathbf f(\mathbf r')|^2\,d^3 r' \notag\\
&=&\int \frac{c_2}{2}\Bigl(\int n^2\,dz'\Bigr)\,|\mathbf f(x',y')|^2\,dx'\,dy' \notag\\
&=&\int \frac{c_2}{2}\,\lambda^2(x',y')\,|\bar{\mathbf M}(x',y')|^2\,dx'\,dy' \notag\\
&\equiv&\int \frac{c_2}{2}\,|\mathbf M_{\rm 2D}(x',y')|^2\,dx'\,dy',
\end{eqnarray}
where we introduce an effective areal magnetization density
\(\mathbf M_{\rm 2D}(x',y')\) with components \(M_{{\rm 2D},i}=\lambda\,\bar M_i\) and
\(\lambda(x',y')=\bigl[\int n^2dz'\bigr]^{1/2}/\bar n\).
To evaluate \(\lambda\), we use the TF approximation, for which
\(\lambda^2=3/(5z'_R)\) and
\(z_R'^2=R_{z'}^2\bigl(1-x'^2/R_{x'}^2-y'^2/R_{y'}^2\bigr)\).
Here \(R_j\) are the TF radii along the $j$ direction and
\(R_{z'}=\bigl[(\omega_{x'}\omega_{y'})/\omega_{z'}^2\,R_{x'}R_{y'}\bigr]^{1/2}\).
The transverse radii \(R_{x',y'}\) are obtained from a TF profile fit to the measured total column density \(\bar n(x',y')\).

We obtain the spin-interaction energy spectrum \(E_i(k)\) from \(M_{{\rm 2D},i}(x',y')\) via Fourier analysis. We select a square region of interest (ROI) of side length
\(L=N_{\rm pix}\Delta r'=160~\mu\mathrm m\) centered on the condensate, where \(N_{\rm pix}\) denotes the number of pixels per side and \(\Delta r'=1.6~\mu\mathrm m\) is the pixel size. In our experiments,
\((R_{x'},R_{y'})\approx(250,130)~\mu\mathrm m\), and the RMS variation of the total column density within the ROI is about \(20\%\).
To suppress spectral leakage due to the finite ROI in the Fourier analysis, we apply a two-dimensional Tukey window with a tapering ratio of \(\mathcal R=0.1\)~\cite{Harris1978Windows}. Explicitly,
\begin{equation}
w_{\mathcal R}(x') =
\begin{cases}
1,
& |x'|<\dfrac{L(1-\mathcal R)}{2},\\[6pt]
\dfrac{1}{2}\left\{1+\cos\!\left[\dfrac{2\pi}{\mathcal R\,L}\!\left(|x'|-\dfrac{L(1-\mathcal R)}{2}\right)\right]\right\},
& \dfrac{L(1-\mathcal R)}{2}\le|x'|\le\dfrac{L}{2},
\end{cases}
\end{equation}
so that the windowed image is expressed as
\(
M^{(0)}_{{\rm 2D},i}(x',y')=w_{\mathcal R}(x')\,w_{\mathcal R}(y')\,M_{{\rm 2D},i}(x',y').
\)

The discrete two-dimensional Fourier transform is given by
\begin{equation}
\widetilde{M}^{(0)}_{{\rm 2D},i}(\mathbf k)
=
\sum_{\mathbf r'_{mn}\in{\rm ROI}}
M^{(0)}_{{\rm 2D},i}(\mathbf r'_{mn})\,
e^{-i\mathbf{k}\cdot\mathbf r'_{mn}}\,\Delta r'^{2},
\label{eq:FT_M}
\end{equation}
where \(\mathbf r'_{mn}=(m\Delta r',n\Delta r')\) and \(\mathbf k=(k_{x'},k_{y'})\).
For an ROI of side length \(L\), the reciprocal-grid spacing is \(\Delta k=2\pi/L\).
The inverse transform reads
\begin{equation}
M^{(0)}_{{\rm 2D},i}(\mathbf r'_{mn})
=
\sum_{\mathbf k}
\widetilde{M}^{(0)}_{{\rm 2D},i}(\mathbf k)\,
e^{+i\mathbf{k}\cdot\mathbf r'_{mn}}\,
\frac{(\Delta k)^2}{(2\pi)^2},
\label{eq:IFT_M}
\end{equation}
and Parseval's theorem gives
\begin{equation}
\sum_{\mathbf r'_{mn}\in{\rm ROI}}
\left|M^{(0)}_{{\rm 2D},i}(\mathbf r'_{mn})\right|^{2}\Delta r'^{2}
=
\sum_{\mathbf k}
\left|\widetilde{M}^{(0)}_{{\rm 2D},i}(\mathbf k)\right|^{2}\,
\frac{(\Delta k)^{2}}{(2\pi)^{2}}.
\end{equation}

We correct the measured power spectrum for the imaging response using the modulation transfer function as
\begin{equation}
\widetilde{M}^{2}_{{\rm 2D},i}(\mathbf{k})
=
\frac{\Bigl\langle
\bigl|\widetilde{M}^{(0)}_{{\rm 2D},i}(\mathbf{k})\bigr|^{2}-B
\Bigr\rangle_{\rm en}}
{|\mathrm{MTF}(\mathbf{k})|^{2}},
\label{eq:MTF_corr}
\end{equation}
where \(B\) is a background offset estimated from the outer border region of the power spectrum and \(\langle\cdots\rangle_{\rm en}\) denotes an ensemble average over 9-15 independent realizations. Within each ensemble, the total atom number was kept within \(10\%\) of the median. In the analysis, Fourier pixels with \(|\mathrm{MTF}(\mathbf k)|^2<0.3\) are excluded.

The one-dimensional spin-interaction energy spectrum for each spin axis \(i\) is obtained by evaluating the spin-interaction energy in an annular bin and dividing by the radial bin width,
\begin{eqnarray}
E_i(k)
&=& \frac{c_{2}}{2}\,\frac{1}{\Delta k}
\int_{||\mathbf{k}'|-k|<\Delta k/2}
\widetilde{M}^{2}_{{\rm 2D},i}(\mathbf{k}')\,\frac{d^{2}\mathbf{k}'}{(2\pi)^2}
\\
&\simeq& \frac{c_{2} k}{4\pi}
\frac{\sum_{\mathbf{k'}}
\omega(\mathbf{k'};k)\,\widetilde{M}^{2}_{{\rm 2D},i}(\mathbf{k'})}
{\sum_{\mathbf{k'}}\omega(\mathbf{k'};k)},
\end{eqnarray}
where the annulus area is approximated by \(2\pi k\,\Delta k\) for \(k\neq0\), and \(\omega(\mathbf{k'};k)\in[0,(\Delta k)^2]\) denotes the overlap area between the Fourier pixel and the annulus. The statistical uncertainty of \(E_i(k)\) is estimated as
\(\sigma_{E_i}(k)=\frac{c_2 k}{4\pi}\sigma_{\rm meas}(k)\), with
\begin{equation}
\sigma^{2}_{\mathrm{meas}}(k)
= \frac{\sum_{\mathbf{k'}}\omega^{2}(\mathbf{k'};k)\,
   \sigma^{2}_{\widetilde{M}^{2}}(\mathbf{k'})}
{\left[\sum_{\mathbf{k'}}\omega(\mathbf{k'};k)\right]^{2}},
\end{equation}
where \(\sigma_{\widetilde{M}^{2}}(\mathbf{k'})\) is the standard error of
\(\widetilde{M}^{2}_{{\rm 2D},i}(\mathbf{k'})\), estimated from the ensemble averaging in Eq.~(\ref{eq:MTF_corr}).

From \(E_i(k)\), the corresponding spin-interaction energy \(\mathcal E_i\) is evaluated as
\begin{equation}
\mathcal{E}_{i}
= \sum_{k\neq 0} E_{i}(k)\,\Delta k +
\frac{c_2 (\Delta k)^2}{32 \pi}\,
\widetilde{M}^{2}_{{\rm 2D},i}(\mathbf{k}=0),
\label{eq:Es_int}
\end{equation}
including the \(k=0\) contribution.
As a reference energy, we compute \(\mathcal{E}_{s0,i}\) for a fully magnetized condensate with the same ROI and Tukey window, using the TF density profile corresponding to the mean atom number \(N\) in the ensemble. The corresponding spin healing length is defined as
\(\xi_{s,i}=\sqrt{\hbar^2/[2m\,(2\mathcal{E}_{s0,i}/N)]}\).

Finally, we construct the total spectrum \(E_s(k)\) by combining the three spectra \(E_i(k)\) (\(i=x,y,z\)). Because the mean atom number differed by approximately \(13\%\) between the measurements of different spin axes, we account for this variation by rescaling each spectrum to common reference values \(\mathcal{E}_{s0}\) and \(\xi_s\) (defined as averages of \(\mathcal{E}_{s0,i}\) and \(\xi_{s,i}\), respectively):
\begin{equation}
E_{s}(k) =
\sum_{i=x,y,z}
\frac{\mathcal{E}_{s0}\xi_{s}}{\mathcal{E}_{s0,i}\xi_{s,i}}\,
E_{i}\!\left(\frac{\xi_{s}}{\xi_{s,i}} k \right).
\label{eq:Es_total_spec}
\end{equation}
Accordingly, the total spin-interaction energy is computed as
\begin{equation}
\mathcal{E}_{s} =
\sum_{i=x,y,z}
\frac{\mathcal{E}_{s0}}{\mathcal{E}_{s0,i}} \,\mathcal{E}_{i}.
\label{eq:Es_total_energy}
\end{equation}
For notational simplicity in the main text, we write
\(E_i(k)/(\mathcal{E}_{s0,i}\xi_{s,i})\) and \(\mathcal{E}_i/\mathcal{E}_{s0,i}\) as
\(E_i(k)/(\mathcal{E}_{s0}\xi_s)\) and \(\mathcal{E}_i/\mathcal{E}_{s0}\), respectively.

\begin{figure*}[h]
  \includegraphics[width=7cm]{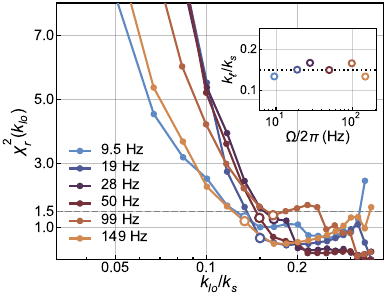}
  \caption{Power-law scaling region. The reduced chi-square \(\chi_r^2(k_{\rm lo})\) was evaluated for weighted power-law fits \(E_s(k)=A k^{-\beta}\) over \(k_{\rm lo}\leq k \leq k_{\rm res}\), where \(k_{\rm res}\) is the resolution-limited maximum wavenumber. Open circles mark the smallest \(k_{\rm lo}=k_\ell\) satisfying \(\chi_r^2<1.5\) (dashed line). Inset: extracted \(k_\ell\) values for different \(\Omega\); the dotted line indicates the mean \(\bar{k}_\ell=0.15(1)k_s\).}
  \label{fig:exponent}
\end{figure*}

\begin{figure*}[h]
  \includegraphics[width=7cm]{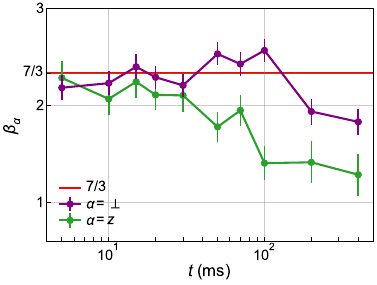}
  \caption{
  Time evolution of the directional power-law exponents during free decay after switching off the RF drive at $\Omega/2\pi=150~\mathrm{Hz}$. The exponents $\beta_\perp(t)$ and $\beta_z(t)$ are extracted from fits to $E_\alpha(k)\propto k^{-\beta_\alpha}$ over $1.3\bar{k}_\ell \le k \le k_{\rm res}$. The red solid line marks $\beta=7/3$. While $\beta_\perp$ remains close to $7/3$ over most of the decay window, $\beta_z$ decreases gradually with time. Error bars denote one standard error.
  }
  \label{fig:sm_decay}
\end{figure*}

\begin{figure*}
  \includegraphics[width=12.27cm]{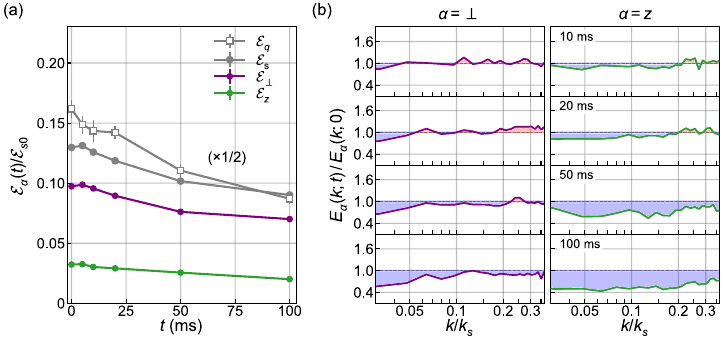}
  \caption{
  Free decay of spin turbulence from a weakly driven steady state at
$\Omega/2\pi=19$~Hz.
(a) Time evolution of the quadratic Zeeman energy $\mathcal{E}_q$ (gray squares), total spin-interaction energy $\mathcal{E}_s$ (gray circles), and its transverse ($\mathcal{E}_\perp=\mathcal{E}_x+\mathcal{E}_y$, purple) and longitudinal ($\mathcal{E}_z$, green) components after the RF drive is switched off. Unlike the case of $\Omega/2\pi=150$~Hz in Fig.~3(a) of the main text, no discernible energy surge is observed in $\mathcal{E}_\perp$; all energy components show an overall monotonic decay as the system relaxes toward the polar ground state. Error bars denote one standard error of the mean.
(b) Spectral ratios $E_{\alpha}(k,t)/E_{\alpha}(k,0)$ for the transverse ($\alpha=\perp$, left) and longitudinal ($\alpha=z$, right) components at different hold times. Red (blue) shading indicates enhancement (depletion) relative to $t=0$.
}
  \label{fig:sm_decay19}
\end{figure*}

\newpage
\subsection{Energy injection rate in steady turbulence}

In this section, we introduce a phenomenological model of how the energy-injection rate in the turbulent steady state depends on the drive frequency \(\Omega\). As discussed in the main text, we assume that the quadratic Zeeman energy \(\mathcal{E}_q\) provides the primary energy reservoir and that energy is transferred into the spin-interaction sector through the buildup of transverse magnetization via spin-mixing dynamics.

Motivated by previous experiments~\cite{hong2023spin,kang2020crossover} and by our own observations, we note that spin populations relax toward the ground-state configuration with an approximately exponential time dependence when a \(\mathcal{E}_q\) value is large. We therefore model the magnetization generated over an effective evolution time \(t\) as proportional to \(1-\exp(-\gamma t)\), where \(\gamma\) is a characteristic growth rate. Assuming that the newly generated magnetization is uncorrelated with the preexisting spin texture, the corresponding increase in spin-interaction energy is taken to scale with the square of this buildup,
\[
\Delta \mathcal{E}_s \;=\; \eta\,\mathcal{E}_q\,[1-\exp(-\gamma t)]^2,
\]
where \(\eta\) is an effective conversion factor.
For a periodically driven steady state, we estimate \(t\) by the drive period \(\tau\simeq 2\pi/\Omega\). Then, the injection rate is suggested as
\begin{equation}
  \Pi(\Omega)=\frac{\Delta \mathcal{E}_s}{\tau}
  \simeq
  \eta\,\mathcal{E}_q(\Omega)\,\frac{\Omega}{2\pi}
  \left[1-\exp\!\left(-\frac{2\pi\gamma}{\Omega}\right)\right]^2 .
  \label{eq:Pi_q_model}
\end{equation}
This expression interpolates between two limits: $\Pi \propto \mathcal{E}_q \gamma^2 / \Omega$ for fast driving ($\Omega/2\pi \gg \gamma$), whereas $\Pi \propto \mathcal{E}_q \Omega$ for slow driving ($\Omega/2\pi \ll \gamma$).

In a statistically steady state, \(\Pi\) may be interpreted as an approximately scale-independent flux of spin-interaction energy through the inertial range, balanced by dissipation at small scales. Hydrodynamic theory with constant flux \(\Pi\) predicts
\(E_s(k)\propto \Pi^{2/3}k^{-7/3}\)~\cite{fujimoto2013spin}. We write this relationship in dimensionless form as
\begin{equation}
  \frac{E_s(k)}{(\mathcal{E}_{s0}/k_s)}
  =
  C\left(
  \frac{\Pi\tau_s}{\mathcal{E}_{s0}}
  \right)^{2/3}
  (k/k_s)^{-7/3},
  \label{eq:cascade_dimless}
\end{equation}
where \(\tau_s=h/\varepsilon_s\simeq 20~\mathrm{ms}\) is the characteristic timescale of the spin dynamics and \(C\) is a dimensionless cascade coefficient assumed to be independent of \(\Omega\).

From the measured spectra, we extract a single prefactor \(A_p\) by fitting the high-\(k\) region to
\(\,E_s(k)/(\mathcal{E}_{s0}/k_s)=A_p(k/k_s)^{-7/3}\).
Comparing with Eq.~(\ref{eq:cascade_dimless}) yields
\(A_p=C\left(\Pi\tau_s/\mathcal{E}_{s0}\right)^{2/3}\),
and combining this with Eq.~(\ref{eq:Pi_q_model}) gives
\begin{equation}
  \frac{A_p^{3/2}}{\tau_s\,(\mathcal{E}_q/\mathcal{E}_{s0})}
  =
  C^{3/2}\eta
  \frac{\Omega}{2\pi}
  \left[
  1-\exp\!\left(
  -\frac{2\pi\gamma}{\Omega}
  \right)
  \right]^2 .
  \label{eq:Ap_model}
\end{equation}

\begin{figure*}[t]
  \includegraphics[width=7.5cm]{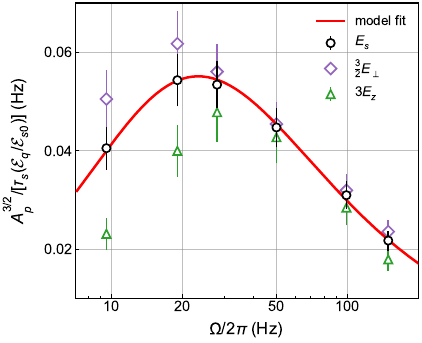}
  \caption{
  Energy injection rates for various \(\Omega\). The prefactor \(A_p\) was extracted by fitting the total spin-interaction energy spectrum to \(E_s(k)/(\mathcal{E}_{s0}/k_s)=A_p (k/k_s)^{-7/3}\) over \(1.3\bar{k}_\ell \le k \le k_{\rm res}\). Here \(\tau_s=h/\varepsilon_s\) is the characteristic timescale of spin dynamics. Error bars indicate \(1\sigma\) uncertainties propagated from the fit uncertainty of \(A_p\) and from the statistical uncertainties in \(\mathcal{E}_q\), \(\mathcal{E}_{s0}\), and \(\tau_s\). The solid red curve is a fit to the model curve, \(A_p^{3/2}/[\tau_s(\mathcal{E}_q/\mathcal{E}_{s0})]=\tilde{\eta}\,\frac{\Omega}{2\pi}\,[1-\exp(-2\pi\gamma/\Omega)]^2\)
[Eq.~(\ref{eq:Ap_model})]. Diamonds and triangles show component-resolved prefactors \(A_p\) extracted from \(\tfrac{3}{2}E_\perp(k)\) and \(3E_z(k)\), respectively.
  }
  \label{fig:collapse}
\end{figure*}

\label{subsec:injection_ansatz}

Figure~\ref{fig:collapse} shows \(A_p^{3/2}/ [\tau_s(\mathcal{E}_q/\mathcal{E}_{s0})]\) as a function of \(\Omega\). Fitting to Eq.~(\ref{eq:Ap_model}) yields \(\gamma=29(2)\) Hz and the combined coefficient \(\tilde{\eta}\equiv C^{3/2}\eta=4.6(5)\times10^{-3}\). The fitted \(\gamma\) is close to the dynamical-instability scale \(\sqrt{q(2c_2n_0-q)}/h\simeq 30\) Hz, which supports interpreting \(\gamma\) as the characteristic magnetization-growth rate sampled within a drive cycle~\cite{bookjans2011quantum, symes2018dynamics}. Using the early-time decay data of \(\mathcal{E}_s\) and \(\mathcal{E}_q\) in Fig.~3(a) of the main text, we estimate an initial transfer ratio \(\eta\simeq 0.25\) over the first \(\sim5\) ms, which implies \(C=(\tilde{\eta}/\eta)^{2/3}\approx 0.07\). While this estimate is necessarily approximate, the model provides a practical guide for parameterizing energy injection and connecting it to the observed spectral amplitude in driven steady states. A more systematic determination of \(\eta\) and \(C\), for example by varying \(q\) and independently characterizing dissipation, is left for future work.


\begin{thebibliography}{99}
\bibitem{footnote3}
The scaling was also found to be robust to the choice of ROI. The residual influence of the smooth density envelope is expected to be concentrated at low wavenumbers, $k\lesssim0.05\,k_s$, well below the power-law fitting range.
\bibitem{footnote2}
$\mathcal{E}_q$ cannot be directly transferred to $\mathcal{E}_z$ due to the conservation of longitudinal magnetization in the absence of spin driving.
\end{thebibliography}
\end{document}